\documentclass{svjour3}                     
\smartqed  
\usepackage{cite}
\usepackage{graphicx}
 \journalname{Gen. Relat. Grav.}
\begin{document}

\title{Thermodynamics and Phase transition of the Reissner-Nordstr\"om Black Hole surrounded by quintessence
}


\author{Bouetou Bouetou Thomas \and
        Mahamat Saleh         \and
        Timoleon Crepin Kofane 
}


\institute{Bouetou B T \and Mahamat Saleh \and
           T C Kofane \at
              Department of Physics, Faculty of Science, University of Yaounde I, P.O. Box. 812, Cameroon \\
           \and
           Bouetou B T
           \at
              Ecole Nationale Sup$\acute{e}$rieure Polytechnique, University of Yaounde I, P.O. Box. 8390,
              Cameroon
              \and
            Mahamat Saleh
             \at
              Department of Physics, Higher Teachers' Training College, University of Maroua, P.O. Box. 55, Cameroon, \\\email{mahsaleh2000@yahoo.fr}}

\date{Received: date / Accepted: date}

\maketitle

\begin{abstract}
We investigate thermodynamics and Phase transition of the
Reissner-Nordstr\"om black hole surrounded by quintessence. Using
thermodynamical laws of black holes, we derive the expressions of
some thermodynamics quantities for the Reissner-Nordstr\"om black
hole surrounded by quintessence. The variations of the temperature
and heat capacity with the entropy were plotted for different values
of the state parameter related to the quintessence, $\omega_{q}$,
and the normalization constant related to the density of
quintessence $c$. We show that when varying the entropy of the black
hole a phase transition is observed in the black hole. Moreover,
when increasing the density of quintessence, the transition point is
shifted to lower entropy and the temperature of the black hole
decreases. 
\end{abstract}

\section{Introduction}
Hawking's discovery of thermal radiation from black
holes was a complete surprise to most specialists, even though quite
a few indications of a close relationship between black hole physics
and thermodynamics had emerged before this discovery\cite{R1}.
Recently, a new idea that gravity can be explained as an entropic
force caused by the information changes when a material body moves
away from the holographic screen was pointed out by
Verlinde\cite{R2}. He showed that through the holographic principle
and the equipartition law of energy that Newton's law of gravitation
can arise naturally and unavoidably in theory in which space is
emergent through a holographic scenario, and a relativistic
generalization leads to the Einstein equation.

One of the most known solutions of Einstein equation is black hole
solution. Black holes are the most fascinating objects in general
relativity. One of important characteristics of a black hole is its
thermodynamical properties. The discovery that black holes laws are
thermodynamical in nature\cite{haw, R3} implies that there should be
an underlying statistical description of them in terms of some
microscopic states. Black hole thermodynamics is widely studied in
the literature\cite{R8, qiyan, amit, qin, gon, yu, R9, R10,R11}.
 It is well known that the heat capacity
of the Schwarzschild black hole is always negative and so the black
hole is thermodynamically unstable. But for the Reissner-Nordstr\"om
black hole, the heat capacity is negative in some parameter region
and positive in other region. Davies pointed out that the phase
transition appears in black hole thermodynamics and the second order
thermodynamic phase transition takes place at the point where the
heat capacity diverges\cite{R4, R5, R6}. Husain and Mann\cite{R7}
suggest that the specific heat of a black hole becomes positive
after a phase transition near the Planck scale. Recently, Jing and
Pan\cite{R8} investigated second order thermodynamic phase
transition for Reissner-Nordstr\"om black hole. Banerjee, Ghosh and
Roychowdhury\cite{Baner} investigated phase transition in
Reissner-Nordstr\"om-AdS black hole and its thermodynamics geometry.

From the recent measurements, we can see that our universe is dominated
by a mysterious form of energy called "Dark Energy". This kind of energy
is responsible of the accelerated expansion of our universe. Today inflation is quite well understood in terms of its
phenomenology, but it still has a number of unresolved foundational
questions. Despite these, it is certainly clear that cosmic
inflation requires a period of cosmic acceleration that cannot be
described by a cosmological constant\cite{rol, cliu}. Cosmic inflation is understood
to be driven by some matter field typically called the "inflaton"
which exhibits an equation of state $p=\rho_q\omega_q$. In this category we can find
quintessence\cite{kis, shuang}, phantom\cite{mart}, k-essence\cite{yang} and quintom\cite{zong, jun}
models.

In this paper, the thermodynamics and phase transition are
investigated for the Reissner-Nordstr\"om black hole surrounded by
quintessence. We consider the metric of a Reissner-Nordstr\"om black
hole surrounded by quintessence\cite{R9, Rx, Rxx, Rxxx} and, from
the laws of black holes thermodynamics, we derived the expressions
of thermodynamics quantities of the black hole. Then we plot the
behavior of heat capacity and temperature versus the entropy for
different values of the density of quintessence in order to extract
the effect of the quintessence on the thermodynamical behavior of
the black hole.

\section{Thermodynamics phase transition of the black hole}

From Kiselev's investigations\cite{kis} on spherically symmetric
solutions for Einstein equations describing black holes surrounded
by quintessence with the energy momentum tensor, which satisfies the
conditions of additivity and linearity, the metric of the
Reissner-Nordstr\"om black hole space-time surrounded by
quintessence can be written as:
\begin{equation}\label{1}
ds^{2}=f(r)dt^{2}-f(r)^{-1}dr^{2}-r^{2}(d\theta^{2}+sin^{2}\theta
d\varphi^{2}),
\end{equation}
where
\begin{equation}\label{2}
    f(r)=1-\frac{2M}{r}+\frac{Q^{2}}{r^{2}}-\frac{c}{r^{3\omega_{q}+1}},
\end{equation}
  $M$ is the black hole mass, $Q$ is the charge of
the black hole, $\omega_{q}$ the quintessential state parameter, $c$
is the normalization factor related to the density of quintessence
$\rho_{q}=-\frac{c}{2}\frac{3\omega_{q}}{r^{3(\omega_{q}+1)}}$.

The event horizon of the black hole can be found from the following equation
\begin{equation}\label{neq}
    f(r)=0.
\end{equation}
This equation will lead us to two event horizons: the outer and inner horizons given by\cite{R9, msh}
\begin{eqnarray}
 \nonumber r_+ &\simeq& M+\sqrt{M^2-Q^2}+\frac{c\Big(M+\sqrt{M^2-Q^2}\Big)^{1-3\omega_q}}{2\sqrt{M^2-Q^2}}, \\
 \nonumber r_- &\simeq& M-\sqrt{M^2-Q^2}-\frac{c\Big(M-\sqrt{M^2-Q^2}\Big)^{1-3\omega_q}}{2\sqrt{M^2-Q^2}},
\end{eqnarray}
respectively.

Taking into account the above equations, the black hole's mass can be expressed as:
\begin{equation}\label{3}
    M=\frac{1}{2}\big[r_++\frac{Q^2}{r_+}-\frac{c}{r^{3\omega_q}_+}\big].
\end{equation}

The entropy of the black hole is given by the area law as:
\begin{equation}\label{4}
    S=\frac{A}{4}=\pi r^2_+.
\end{equation}

\begin{figure}[h!]
\begin{center}
  \includegraphics[width=10cm]{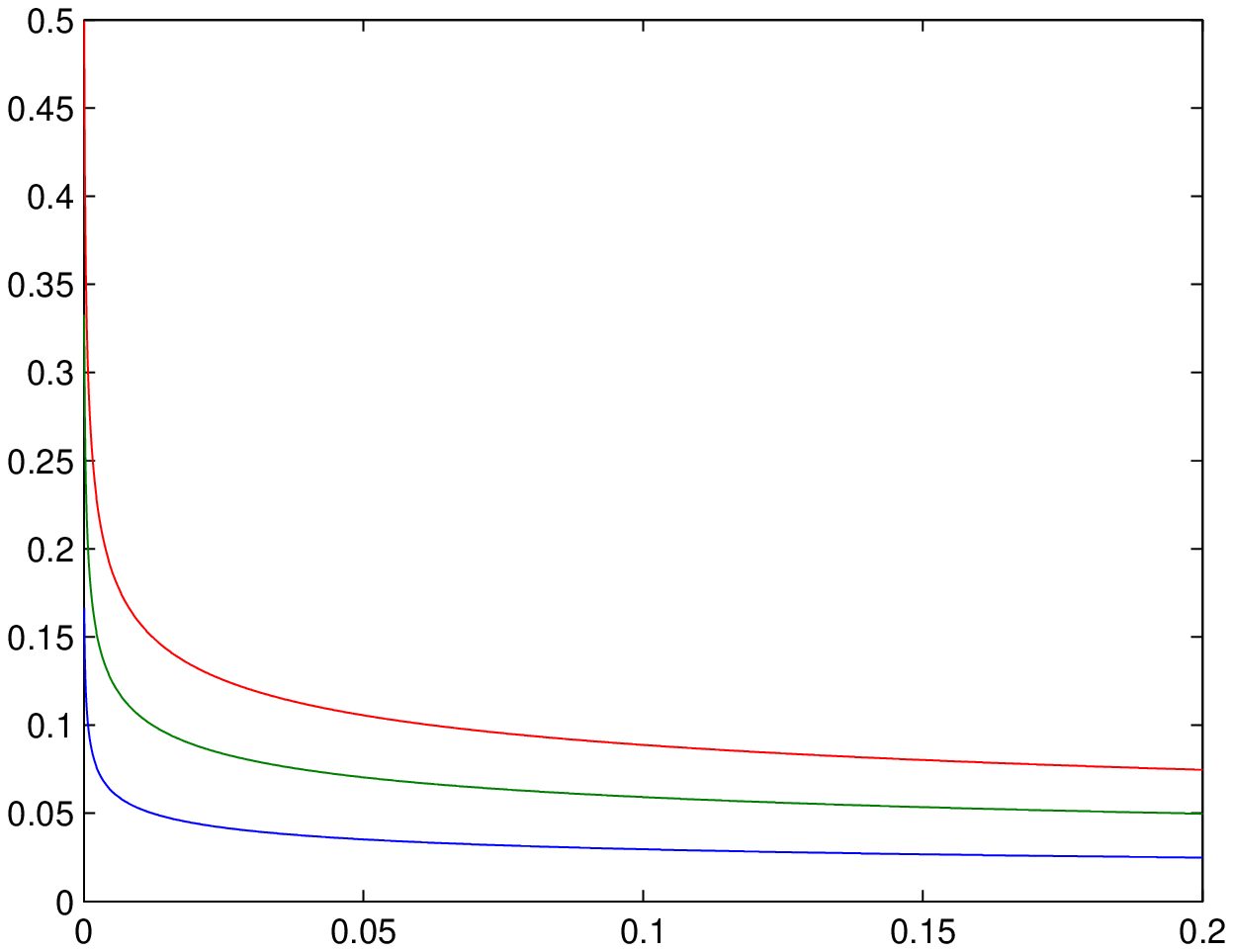}\\
  \caption{Variation of the density of quintessence versus entropy for different values of $c$}\label{dc}
  \end{center}
\end{figure}

\begin{figure}[h!]
\begin{center}
  \includegraphics[width=10cm]{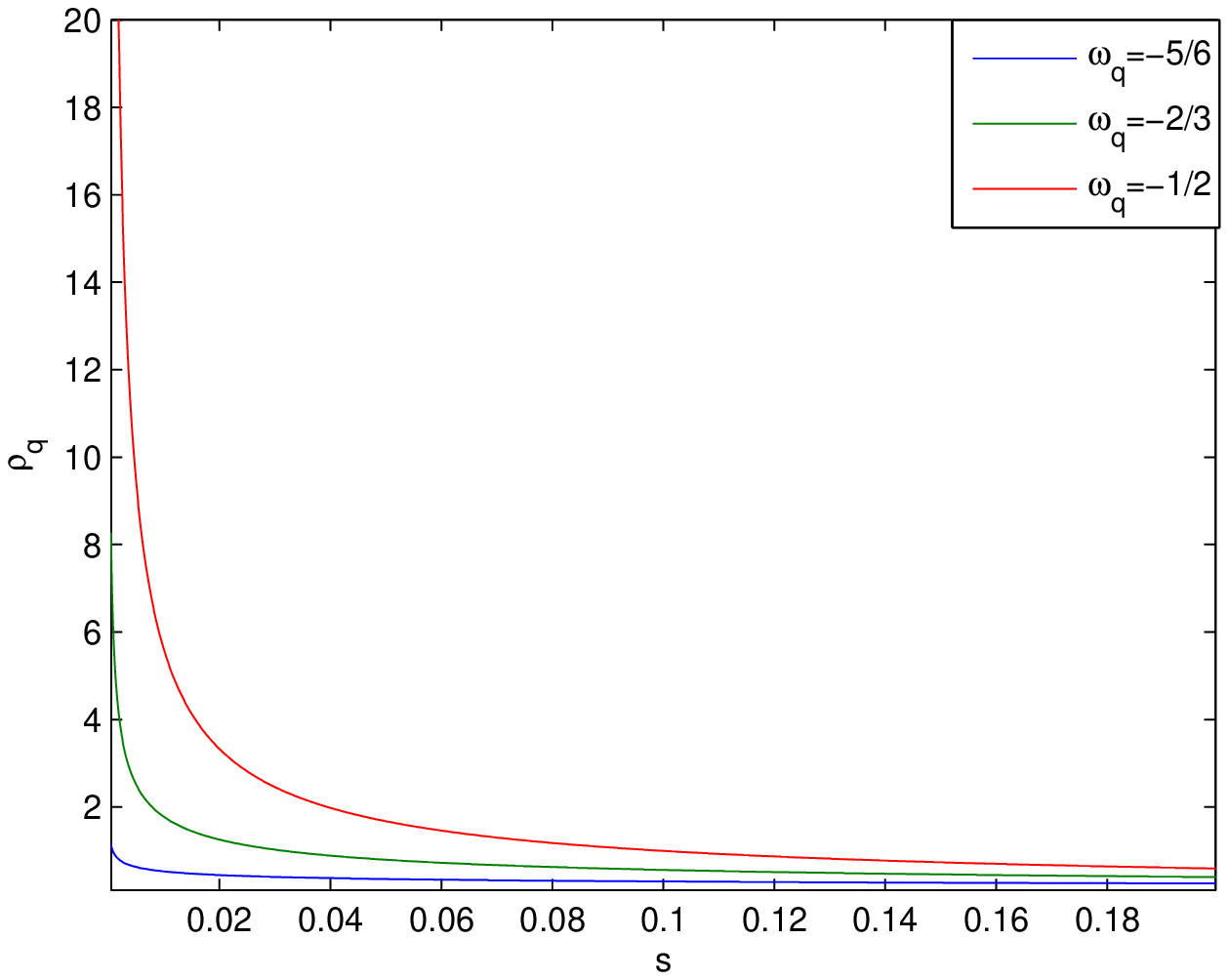}\\
  \caption{Variation of the density of quintessence versus entropy for different values of $\omega_q$}\label{dw}
\end{center}
\end{figure}

Substituting (4) into (3), the black hole mass can be rewritten as a function of the black hole entropy
\begin{equation}\label{5}
    M=\frac{1}{2}\Big[\sqrt{\frac{S}{\pi}}+Q^2\sqrt{\frac{\pi}{S}}-c\Big(\frac{\pi}{S}\Big)^{\frac{3\omega_q}{2}}\Big].
\end{equation}
With the same considerations, the density of quintessence at the event horizon of the black hole can be expressed in function of the entropy as:
\begin{equation}\label{d}
    \rho_q=-\frac{3c\omega_q}{2}\Big(\frac{\pi}{S}\Big)^{\frac{3\omega_q+3}{2}}
\end{equation}
Its behavior is plotted in Figs. \ref{dc} and \ref{dw}

Using the first law of black hole thermodynamics for the
Reissner-Nordstr\"om black hole\cite{R1}
\begin{equation}\label{6}
    \delta M=T\delta S+\phi\delta Q,
\end{equation}
it is possible to define the other thermodynamics quantities. These quantities are given by:
\begin{eqnarray}
  T&=&\Big(\frac{\partial M}{\partial S}\Big)_Q =\frac{1}{4(\pi S)^{\frac{3}{2}}}\Big(-\pi^2Q^2+\pi S +3c\omega_q\pi^{\frac{3\omega_q+3}{2}}S^{\frac{1-3\omega_q}{2}}\Big) \\
  \phi&=&\Big(\frac{\partial M}{\partial Q}\Big)_S = Q\sqrt{\frac{\pi}{S}} \\
  C_\phi&=&T\Big(\frac{\partial S}{\partial T}\Big)_\phi=2S\frac{-\pi Q^2+S+3c\omega_q\pi(\frac{\pi}{S})^{\frac{3\omega_q-1}{2}}}{\pi
  Q^2-S-6c\omega_q\pi(\frac{\pi}{S})^{\frac{3\omega_q-1}{2}}} \\
  C_Q &=&T\Big(\frac{\partial S}{\partial T}\Big)_Q=-2S\frac{S-\pi Q^2+3c\omega_q\pi^{\frac{1+3\omega_q}{2}}S^{\frac{1-3\omega_q}{2}}}{S-3\pi Q^2+3c(2+3\omega_q)\pi^{\frac{1+3\omega_q}{2}}S^{\frac{1-3\omega_q}{2}}}
\end{eqnarray}
where
$\phi$ represents the potential difference between the horizon and
infinity, $T$ is the Hawking temperature of the black hole,
$C_\phi$ is the heat capacity at constant potential of the black hole and
$C_Q$ is the heat capacity at constant charge of the black hole.

In the absence of quintessence ($c\omega_q=0$), the heat capacity transforms to
\begin{equation}\label{ne}
    C_\phi=-2S<0.
\end{equation}
This expression is identic to that of the Schwarzschild black hole. In this situation, the only difference between the heat capacities of Schwarzschild and Reissner-Nordstr\"om black holes come from the expression of the entropy which depends on the black hole parameters (mass and charge).
It means that the Reissner-Nordstr\"om black hole is also thermodynamically  unstable in the absence of quintessence although there is an infinite discontinuity in the heat capacity at constant charge $C_Q$. In fact, Hut\cite{hut} argued that although the finite discontinuity in $C_Q$ can be classified as a first order phase transition, the infinite discontinuity cannot be considered as a phase change at all. But this infinite discontinuity has a physical significance which in a way transcends that of phase transition. Although this does not affect the internal state of the system as in the case of a phase transition, it indicates transition from a region where only a microcanonical ensemble is appropriate to a region where a canonical ensemble too can be used to describe the system. Now on the phase transition can be observed through the behavior of $C_\phi$\cite{Baner}.

\begin{figure}[h!]
\begin{center}
\includegraphics[width=10cm, height=8cm]{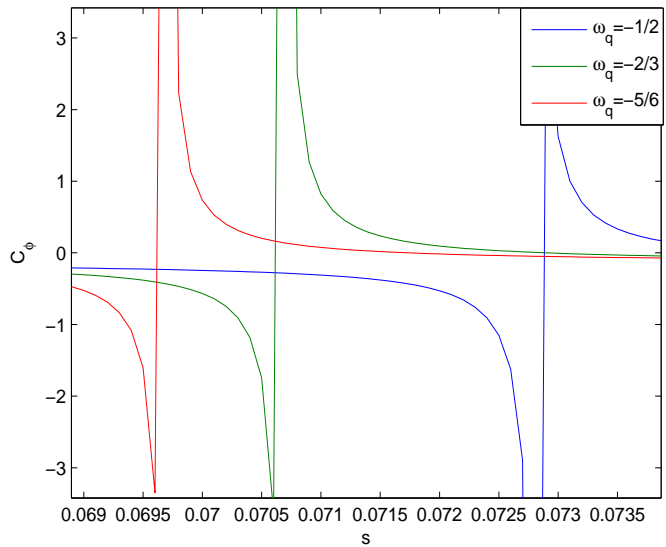}
\caption{Variation of the heat capacity at constant potential versus entropy for different
values of the state parameter $\omega_{q}$.}\label{fig1}
\end{center}
\end{figure}

\begin{figure}[h!]
\begin{center}
\includegraphics[width=10cm, height=8cm]{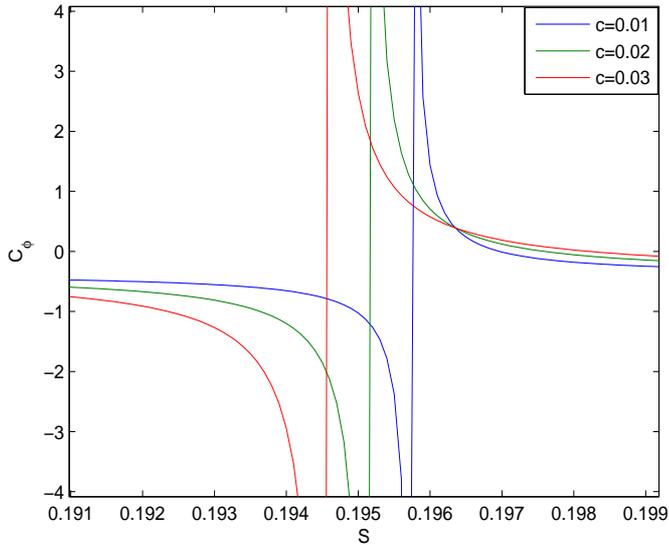}
\caption{Variation of the heat capacity at constant potential versus entropy for different
values of c.}\label{fig2}
\end{center}
\end{figure}

Explicitly, we plot the variation of the heat capacity $C_\phi$ versus the
entropy for different values of the state parameter of the
quintessence $\omega_q$ and the normalization constant $c$,
respectively. Its behavior is represented in Fig.~\ref{fig1} and
Fig.~\ref{fig2}, respectively.

 Through these figures, we can remark that when varying the entropy,
the heat capacity passes a discontinuity point representing the phase transition point of the black hole. Moreover, this point
is shifted to lower entropy when decreasing $c$ or $\omega_q$.

\begin{figure}[h!]
\begin{center}
\includegraphics[width=10cm, height=8cm]{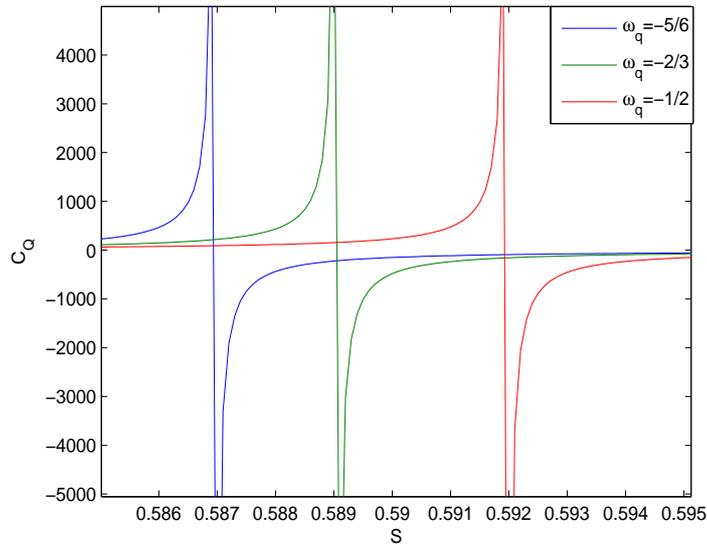}
\caption{Variation of the heat capacity at constant charge versus entropy for different
values of the state parameter $\omega_{q}$.}\label{hcapq1}
\end{center}
\end{figure}

\begin{figure}[h!]
\begin{center}
\includegraphics[width=10cm, height=8cm]{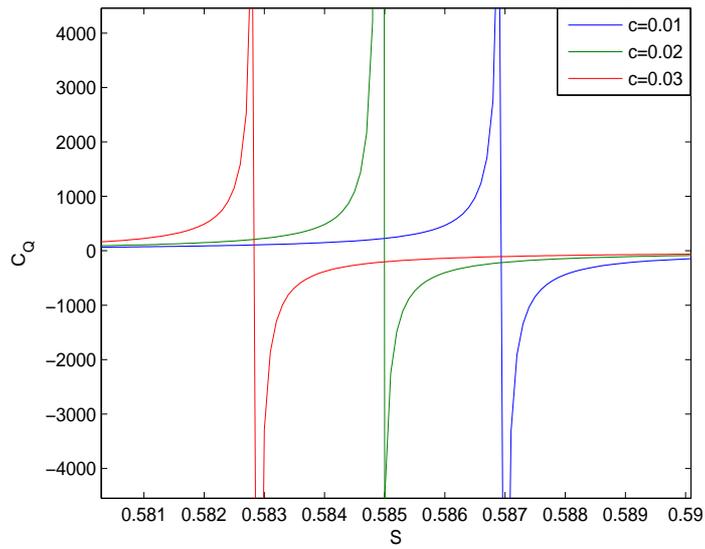}
\caption{Variation of the heat capacity at constant charge versus entropy for different
values of $c$.}\label{hcapq2}
\end{center}
\end{figure}

We also plot the variation of the heat capacity at constant charge $C_Q$ versus the
entropy for different values of the state parameter of the
quintessence $\omega_q$ and the normalization constant $c$,
respectively. Its behavior is represented in Fig.~\ref{hcapq1} and
Fig.~\ref{hcapq2}, respectively.

As for the heat capacity at constant potential $C_\phi$, we can remark through these figures that when varying the entropy,
the heat capacity at constant charge passes a discontinuity point representing the transition from a region where only a microcanonical ensemble is appropriate to a region where a canonical ensemble too can be used to describe the black hole. Moreover, this point
is also shifted to lower entropy when decreasing $c$ or $\omega_q$.

Comparatively, we can remark that for a given black hole, the discontinuity point of $C_\phi$ occurs before that of $C_Q$ when increasing the entropy of the black hole. We can also remark that $C_Q$ is slightly greater than $C_\phi$ ($C_\phi/C_Q\simeq 1.4\times10^{-4}$).

Let us now focus on the temperature. To see its behavior, we plot
its variation versus the entropy for different values of the state
parameter of quintessence, $\omega_q$, and the normalization
constant, $c$, respectively.

\begin{figure}[h!]
\begin{center}
\includegraphics[width=10cm, height=8cm]{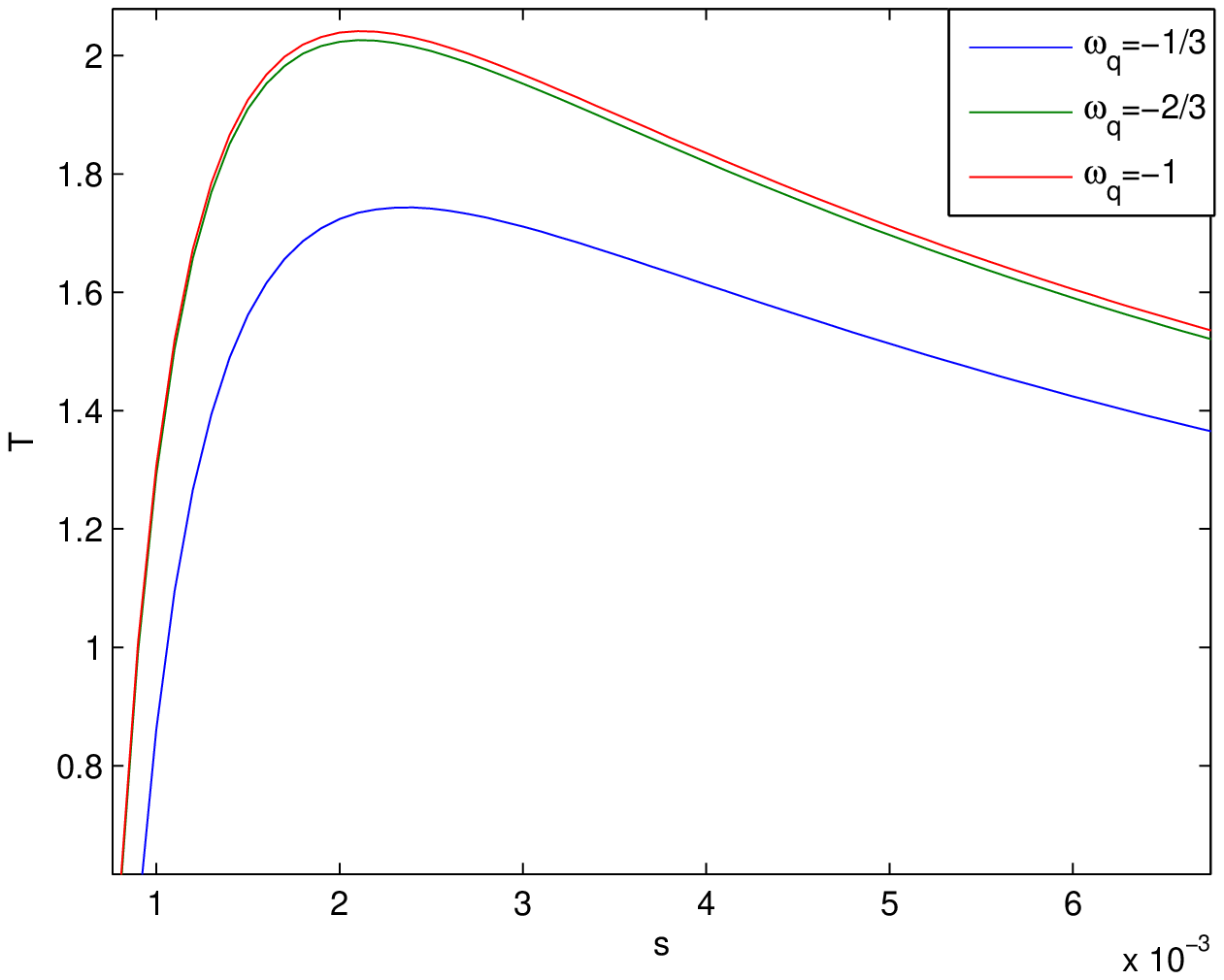}
\caption{Variation of the temperature versus entropy for different
values of the state parameter $\omega_{q}$.}\label{fig3}
\end{center}
\end{figure}

\begin{figure}[h!]
\begin{center}
\includegraphics[width=10cm, height=8cm]{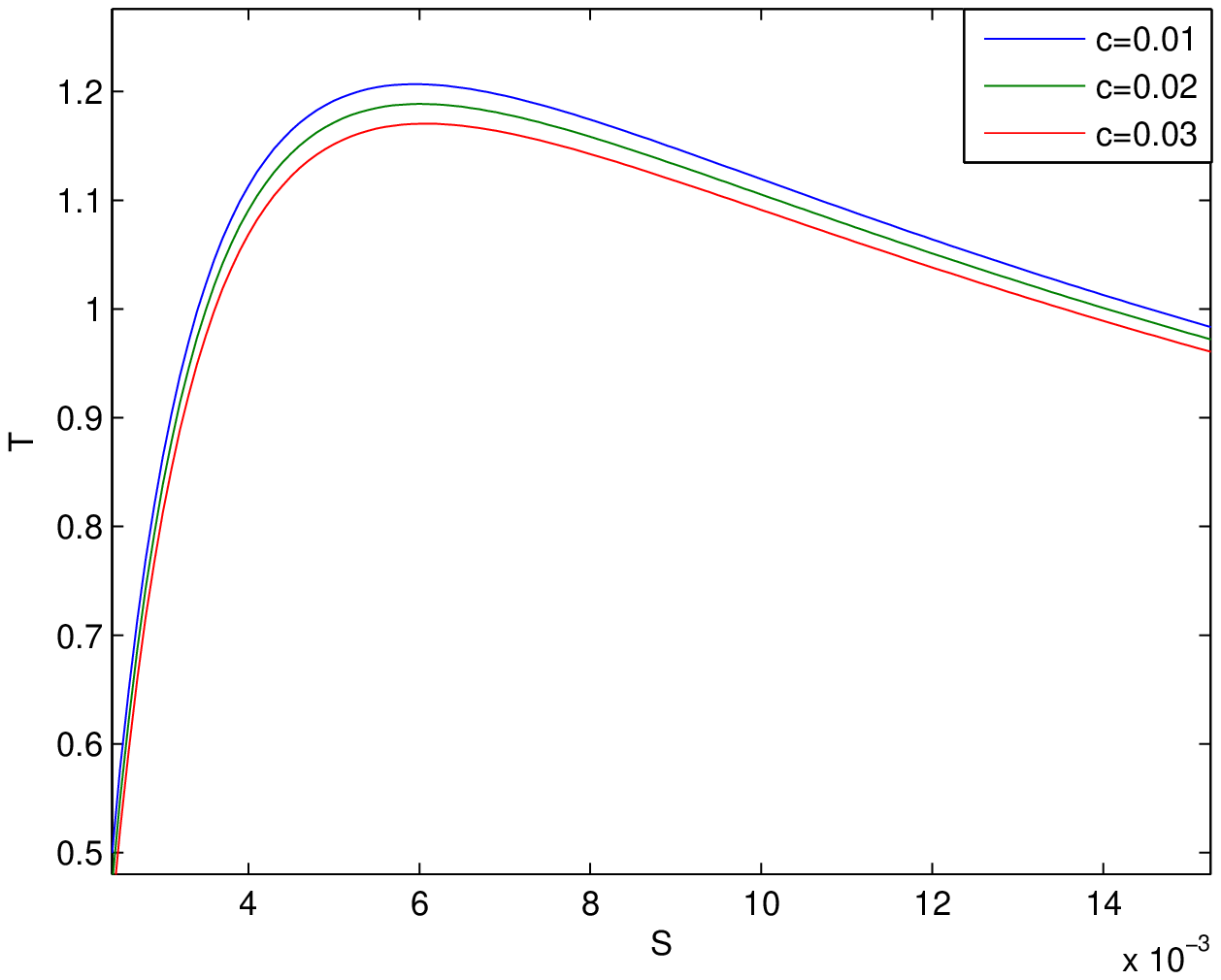}
\caption{Variation of the temperature versus entropy for different
values of c.}\label{fig4}
\end{center}
\end{figure}

From these figures, we can see that there is no discontinuity of
temperature when varying the entropy $S$. We can also remark that
the temperature decreases when increasing $c$ or $\omega_q$.

\section{Summary and Conclusion}

Quintessence surrounding the Reissner-Nordstr\"om black hole
modifies the background metric of the black hole. This modification
is pointed out by the last term of Eq.\ref{2}. This term is just a
perturbation term since the normalization constant related to the
density of quintessence is actually more smaller than
$0.001$\cite{zhangy}. From the figures plotted above, we can see
that the phase transition of the black hole is shifted to lower
entropy when increasing $c$ or $\omega_q$. Moreover, the temperature
of the black hole is decreasing when increasing $c$ or $\omega_q$.
From Figs. \ref{dc} and \ref{dw} we can see that in this scale of
entropy, increasing $c$ or $\omega_q$ means increasing the density
of quintessence. Thus we can conclude that when increasing the
density of quintessence surrounding the Reissner-Nordstr\"om black
hole, the phase transition point of the black hole is shifted to
lower entropy. We can also conclude that quintessence acts as a cold
field decreasing the temperature of the black hole as stepped
previously\cite{R9}. Moreover, we remark that when the density of quintessence vanishes, the black hole becomes thermodynamically unstable and there is no phase transition observed since the heat capacity becomes continuously negative. We can conclude that quintessence helps stabilizing thermodynamically the Reissner-Nordstr\"om black hole. From Fig.\ref{fig3}, we can see that the variation of the temperature becomes more and more negligible when approaching the phantom limit ($\omega_q=-1$). It will be then interesting to investigate thermodynamical behavior of the Reissner-Nordstr\"om black hole in the presence of quintom dark energy in future works.

\end{document}